# Ultrafast switchable spin-orbit coupling for silicon spin qubits via spin valves

Ranran Cai[1,2,†], Fang-Ge Li[1,2,†], Bao-Chuan Wang[1,2], Hai-Ou Li[1,2,3], Gang Cao[1,2,3,*] and Guo-Ping Guo[1,2,3,4,*]

[1]CAS Key Laboratory of Quantum Information, University of Science and Technology of China, Hefei, Anhui 230026, P. R. China

[2]CAS Center for Excellence in Quantum Information and Quantum Physics, University of Science and Technology of China, Hefei, Anhui 230026, China

[3]Hefei National Laboratory, University of Science and Technology of China, Hefei 230088, China

[4]Origin Quantum Computing Company Limited, Hefei, Anhui 230088, China

[†]These authors contributed equally to the work

*Correspondence to: gcao@ustc.edu.cn (G.C.) and gpguo@ustc.edu.cn (G.G.)

**Recent experimental breakthroughs, particularly for single-qubit and two-qubit gates exceeding the error correction threshold, highlight silicon spin qubits as leading candidates for fault-tolerant quantum computation. In the existing architecture, intrinsic or synthetic spin-orbit coupling (SOC) is critical in various aspects, including electrical control, addressability, scalability, *etc*. However, the high-fidelity SWAP operation and quantum state transfer (QST) between spin qubits, crucial for qubit-qubit connectivity, require the switchable nature of SOC which is rarely considered. Here, we propose a flexible architecture based on spin valves by electrically changing its magnetization orientation within sub-nanoseconds to generate ultrafast switchable SOC. Based on the switchable SOC architecture, both SWAP operation of neighbor spin qubits and resonant QST between distant spins can be realized with fidelity exceeding 99% while considering the realistic experimental parameters. Benefiting from the compatible processes with the modern**

**semiconductor industry and experimental advances in spin valves and spin qubits, our results pave the way for future construction of silicon-based quantum chips.**

**Introduction**

Spin qubit confined in silicon-based quantum dots (QDs)[1] have attracted extensive attention due to its long coherence time[2] and compatibility with modern semiconductor industry[3,4]. Spin-orbit coupling (SOC), whether intrinsic or synthetic from stray fields, plays a critical role in various aspects of silicon spin qubits[5], including electrical control, addressability and scalability *etc*. It has achieved significant successes in single-qubit[6,7] and two-qubit (CZ, CROT) gates[8-10] with fidelity exceeding 99%, making it a competitive route to realize the fault-tolerant quantum computing[11]. With such ambition, the scaling-up of spin qubits is essential, which requires not only neighbor connections via exchange coupling[12] but also remote coupling[13,14]. Circuit quantum electrodynamics (cQED)[15], where the microwave photon serves as quantum information medium, provides an effective on-chip scheme for long-distance coupling. It has recently been realized in silicon QD-cavity hybridization system[16,17] with the synthetic SOC from micromagnet induced[18-21].

However, SOC tend to expose spin qubit to charge noise to reduce its coherence. Moreover, there are key issues also rooted in SOC waiting to be addressed for further coherent quantum information exchange between spin qubits, no matter local or remote. Specifically, the inevitable energy shift between local spin qubits arising from the SOC or stray field restricts the fidelity of two-qubit SWAP operation[22-24]. In addition, although the virtual photon mediated *i*SWAP is repoted[25], the resonant quantum state transfer (QST), widely used for distributed quantum processors and networks[26-31], has not been explored in silicon spin qubits because the SOC-induced spin qubit-microwave photon coupling is difficult to quickly switch on/off [20,32]. To eliminate these limitations and without loss of other outstanding achievements, the SOC should be switched on/off at proper times[7] which is rarely discussed in previous investigations that refer to static SOC architectures.

Spin valves, consisting of two magnetic layers separated by a tunneling barrier, draw extensive interest in the field of spintronics[33] where mainly focused on magnetic storage for commercial magnetic-random-access-memory (MRAM) applications[34]. Recently, the synthetic SOC from stray field, inherent to the magnetization configuration of spin valves, has been recognized and applied to manipulate Majorana bound state (MBS)[35,36]. Strikingly, since its magnetization configuration can be electrically switched within sub-nanoseconds[37,38], the synthetic SOC is thus ultrafast switchable, which should be excellently meet the requirement of silicon spin qubits.

In this paper, we first utilize spin valves to construct switchable SOC. Subsequently, both the SWAP operation of neighbor spins and resonant QST between distant spins have been proven to be realized with fidelity exceeding 99% by considering realistic experimental parameters based on our switchable SOC architecture.

**Switchable SOC based on spin valves**

First, we start with the discussion of how to achieve switchable SOC through spin valves in detail. As shown in Fig. 1**a**, an electron spin qubit is confined in silicon-based double quantum dot (DQD). Its energy detuning ($\varepsilon$) and tunnel coupling ($2t_\text{c}$) are controlled by plunger (PL and PR) and barrier (BG) gates (red line in Fig. 1**a**), respectively. The applied magnetic field along the $z$ direction induces Zeeman splitting to encode quantum information. Different from the static SOC scheme, micromagnets are replaced with spin valves[33,34] to generate inhomogeneous stray field. To hybridize the spin with charge, the oscillating electron in the DQD experiences stray field gradient along *x* direction acting as SOC. Since the spin valve stray field is inherent to its magnetic configuration, the corresponding SOC is thus highly controllable (Fig. S1). To manipulate SOC, we are interested in spin valve with magnetic tunneling junction (MTJ) geometry which consists of magnetic pinning and free layer separated by oxide barrier (inset in Fig.1**a**). By applying charge current, its magnetization configurations are switched between parallel (P) and antiparallel (AP) via spin polarization from pinning layer and/or spin Hall effect in heavy metals, named spin-transfer torque (STT)[39] and spin-orbit torque (SOT)[40], respectively. Combined with recent experimental advances in sub-nanosecond switching [37,38], spin valve will serve as ultrafast switcher for SOC.

Next, the synthetic SOC is simplified to the difference of *x*-axis stray field between right and left QDs ($b_x = |B_{x,R} - B_{x,L}|$) which is widely considered for SWAP operation[22-24] and spin-photon coupling[20,41]. To simulate the stray field, two spin valves with opposite pinning and free layer sequences are integrated with silicon DQD (Fig. 1**a**). We select CoFe as the magnetic layers, due to its large saturation magnetization ($M_s = 1.7 \times 10^6$ A/m) to guarantee enough stray field even though the spin valve is of small size for ultrafast switching. Figure 1**c** shows *x*-axis stray field for both spin valves in parallel state (P-P) (simulation details in Supplementary Note 1). Since the magnetization is arranged along the same direction, P-P state is consistent with micromagnet, resulting in inhomogeneous stray field spread throughout entire DQD region and the *x*-direction component is opposite for the left ($B_{x,L}$) and right ($B_{x,R}$) QDs (Fig. 1**c**). From the line cut along *y* = 0 (purple line in Fig. 1**c**), the $b_x$ is determined to be ~ 52 mT which is responsible for switching on the SOC[42-44]. However, for AP-AP state, the stray field is localized around spin valves (Fig. 1**d**). More importantly, even parity of *x*-axis stray field with respect to DQD makes the $b_x$ strictly equal to zero to switch off synthetic SOC.

**High-fidelity SWAP operation between local silicon spin qubits**

With the establishment of switchable stray field, it is convenient to discuss the high-fidelity SWAP operation based on our spin valve scheme. Among the three basic two-qubit gates, although the CPHASE and CROT operations have exceeded the error correction threshold[8-10], exploring high-fidelity SWAP operation is still crucial for qubit-qubit connectivity while reducing circuit depth[45,46]. However, the fidelity of SWAP operation is restricted by the requirement of large $\frac{J}{\Delta E_z}$ ratio[22,45], where $J$ is the exchange coupling and $\Delta E_z$ is Zeeman energy difference between two spin qubits. In static SOC architecture, neither intrinsic SOC induced anisotropic *g*-factor nor stray field gradient from micromagnet will induce a considerable $\Delta E_z$ to make the SWAP rotate with a tilted angle[47]. As shown in Fig. 2**a**, the SWAP fidelity rapidly decreases as $\Delta E_z$ increases, especially for smaller $J$, which results in the high-fidelity SWAP being limited in narrow region for $\Delta E_z$ approaching zero (white line in Fig.2**a**), so that is difficult to realize in static SOC architecture[23,24]. Recent efforts have attempted to improve SWAP operation, including energy compensation[23], phase correction[23,24] and high $\frac{J}{\Delta E_z}$ ratio[48], which make the optimal fidelity of *coherent*-SWAP reach 84%. Principally, $\Delta E_z$ should be zero to guarantee perfect SWAP without phase impacts[47]. This can be natural realized in our spin valve architecture where the magnetic field difference will be strictly switched off for AP-AP state (Fig. 1**d**). In Fig. 2**b**, we simulate the SWAP fidelity as a function of $J$ and qubit decoherence rate ($\gamma_\varphi$) by placing spin valves in AP-AP state (detailed calculation in Supplementary Note 2). Such fidelity can surpass 99% for $\gamma_\varphi$ approximately at 0.1 MHz and $J$ at dozens of MHz (white line in Fig. 2**b**), which are typical parameters in practical devices[2]. In addition, since the switching time ($t_{switch}$) is within sub-nanoseconds, after SWAP operation, the spin valves can be quickly switched back to P-P state in order to open synthetic SOC without giving up other functions of silicon spin qubits, including electrical control[7], CPHASE operation[24], *etc*.

**Hamiltonian of DQD-cavity hybridization system**

After the discussion of high-fidelity SWAP operation between local spins, we are interested in the resonant QST between remote spin qubits for long-distance qubit-qubit connectivity. Here, we focus on microwave photons as information medium which is usually explored in DQD-cavity hybridization devices[16,17,42,43] with spin-photon coupling. To model this hybridization system in our switchable SOC architecture, we first consider the Hamiltonian ($H$) of flopping mode in DQD accompanied by spin valves which can be divided into two parts as follows:

$$H = H_{DQD} + H(t), \qquad (1)$$

where $H_{DQD} = \frac{1}{2}(\varepsilon \tau_z + 2t_c \tau_x + B_z \sigma_z + b_x \sigma_x \tau_z)$ is used to describe charge and spin components in DQD with $\tau_k$ and $\sigma_k$, for $k = x, y, z$, being the charge and spin Pauli matrices respectively. The first and second terms of $H_{DQD}$ represent energy detuning ($\varepsilon$) and tunnel coupling ($2t_c$), respectively. $B_z \sigma_z$ is the Zeeman splitting and $b_x \sigma_x \tau_z$ represents the spin-charge hybridization via stray field induced SOC. Thus, spin states $|\uparrow\rangle$ and $|\downarrow\rangle$ are accompanied by bonding ($|+\rangle$)/antibonding ($|-\rangle$) state of charge component to form four eigenenergy levels as shown in Fig. 1**b**. The ground-state $|0\rangle = |-, \downarrow\rangle$ and first excited-state $|1\rangle = |-, \uparrow\rangle$ are chosen to encode quantum information with $2t_c > B_z$ to maintain primarily spin-like character [20]. As the spin valve magnetization configuration changes from P-P to AP-AP states leading to the magnetic field variation, $H(t)$ can be expressed as:

$$H(t) = \frac{1}{2}(B_z(t)\sigma_z + b_x(t)\sigma_x\tau_z). \tag{2}$$

Then, the effects of microwave photon and interaction are introduced to describe the hybridization system. $H_r = \omega_c a^\dagger a$ is the Hamiltonian of the microwave photon with frequency $\omega_c$ which is set to be resonant with spin qubits for resonant QST, where $a$ and $a^\dagger$ denote the annihilation and creation operators. The DQD-cavity interaction is described by $H_I = g_c(a + a^\dagger)\tau_z$, in which $g_c$ is the electric dipole coupling between charge and photon. Due to the SOC induced by transverse field gradient $b_x$, the indirect coupling between the spin state and microwave photon ($g_s$, yellow arrows in Fig. 1**b**) will be achieved and it can be tuned on/off by switching the spin valves. Following, we diagonalize $H_{DQD} + H_r$ at sweet spot ($\varepsilon = 0$) to minimize charge noise and maintain long qubit decoherence time $(T_2^*)$[49]. As such, system Hamiltonian $H_{sys} = H_{DQD} + H_r + H_I + H(t)$ will be rewritten on $H_{DQD} + H_r$ eigen-basis and we can directly calculate the state evolution of spin and photon.

To capture qubit decoherence and photon loss, we use the quantum master equation to model the system,

$$\dot{\rho} = -i[H_{sys}, \rho] + \sum_{i=1}^{2}\left(\gamma_l^i \mathcal{D}[\sigma_-^i]\rho + \frac{\gamma_\phi^i}{2}\mathcal{D}[\sigma_z^i]\rho\right) + \kappa \mathcal{D}[a]\rho, \tag{3}$$

$\rho$ is the density matrix of the two spin qubits and photon. $\mathcal{D}[O]\rho = O\rho O^\dagger - (O^\dagger O \rho + \rho O^\dagger O)/2$ is the Lindblad operator. $\kappa$ is the photon loss rate in resonator. $\gamma_l^i$ and $\gamma_\phi^i$ represent the relaxation rate and pure dephasing rate of each spin qubit, respectively, where $\gamma_l^i \ll \gamma_\phi^i$ in silicon quantum dots [6,42]. Here, $\gamma_\phi^i$ of the silicon spin qubit is mainly inherited from hybridization with charge, neglecting the influence of nuclear spins and other pathways[20].

**Quantum state transfer between remote silicon spin qubits**

Based on the model of DQD-cavity hybridization system, the resonant QST in our spin valve (SV) architecture will be simulated. Two DQDs accompanied by spin valves are capacitively coupled to the end of a coplanar waveguide (CPW) superconducting resonator (Fig. 3**a**) with pulse sequences in Fig. 3**b** where spin-photon coupling ($g_{s1}$, $g_{s2}$) are controlled by spin valves ($SV_{1L,R}$ and $SV_{2L,R}$ in Fig. 3**a**). Initially, two spin qubits and microwave photon frequency are set to resonant at 5.8 GHz and $SV_{1L,R}$ ($SV_{2L,R}$) are in AP-AP state to turn off $g_{s1}$ ($g_{s2}$). After spin qubit-1 (Q1) is prepared into excited state, $SV_{1L,R}$ are switched to P-P configuration to open spin-photon coupling with $g_{s1}$ = 34 MHz, a typical value in experiments[17] (other parameters are summarized in Supplementary Table). To suppress the loss of quantum information during switching process, the switching time is chosen as fast as $t_{switch}$ = 0.2 ns which is an optimal metric for spin valves[37]. After quantum state evolves from Q1 to photon during duration time $T = \frac{\pi}{2g_{s1}}$, $SV_{1L,R}$ are turned off and the coupling between spin qubit-2 (Q2) and photon is opened via $SV_{2L,R}$ to transfer state from photon to Q2. Figure 3**d** shows the excited state probability for each spin qubit and photon during QST process. The fidelity of this process is calculated with $\mathcal{F}_s = \langle\psi|\rho|\psi\rangle$, where $\rho$ is the final state simulated by master equation and $\psi = |g0e\rangle$ is the ideal final state. With the realistic experimental parameters[16,17,42,43] $\gamma_\phi/2\pi = 1$ MHz and $\kappa/2\pi = 1.3$ MHz, the QST fidelity can be achieved to be 91.5% as shown in Fig. 3**g**.

For comparison, in static SOC architecture, $g_{s1}$ ($g_{s2}$) modulation relies on the barrier and plunger gates[50]. As illustrated in Fig. 3**c**, the electron can be pushed into right or central single QD by tuning $\varepsilon$ or $2t_c$ to reduce its experience of stray field difference. To prove the superiority of our spin valve scheme, we simulate the $\varepsilon$ and $2t_c$ controlled QST with other parameters identical to spin valve method. Since it is difficult to fully close $g_{s1}$ ($g_{s2}$) in conventional hybridization devices[32], the gate manipulate $\varepsilon$ and $2t_c$ vary up to 100 GHz to switch off as much as possible. Clearly, a quantum state 'jump' is observed in Fig. 3**e,f** during the $g_{s1}$ ($g_{s2}$) is switched, which suppresses the fidelity to 42.0% (Fig. 3**h**) and 89.0% (Fig. 3**i**) for $\varepsilon$ and $2t_c$ methods, respectively.

To determine the origin of 'jump', we calculate the $\varepsilon$ and $2t_c$ dependent energy levels as shown in Fig. 4**a** and 4**b**, respectively. When changing $\varepsilon$ or $2t_c$ to switch $g_{s1}/g_{s2}$, diabatic process induces significant Landau-Zener (LZ) transition [51] (purple arrow) owing to the rapid energy gap variation within $t_{switch}$. The LZ transition causes quantum information leakage from coding states to higher-excited states, leading to the 'jump' behavior in Fig. 3**e,f**. Since the reduction of detuning or tunnel coupling variation ($\Delta\varepsilon$ and $\Delta 2t_c$) or prolonging $t_{switch}$ can weaken the effect of LZ transition, we simulate the QST fidelity phase diagram as a function of $\Delta\varepsilon$ ($\Delta 2t_c$) and $t_{switch}$ as shown in Fig. 4**d** (4**e**). However, in either two methods, the maximum fidelity is still less than 90% (Fig. S4) and appears in the intermediate region. As $\Delta\varepsilon$ ($\Delta 2t_c$) is further decreased or $t_{switch}$ is broadened, the residual coupling and long interaction time induced state backflow and dephase also suppress fidelity. In contrast, the spin valves primarily modulate the hybridization between spin and charge components, resulting in minimal adjustments to the qubit energy and satisfying the adiabatic approximation even within sub-nanosecond switching (Fig. 4**c**). Therefore, the SV method controlled QST can easily exceed its performance in static SOC architecture. As shown in Fig. 4**f**, the fidelity breaks 90% threshold when $t_{switch}$ = 0.5 ns and continuously increases, indicating that the QST can be further improved with the advances of spin valves by shortening its switching time.

Next, we discuss the possible restriction and corresponding optimization schemes to make QST fidelity exceed 99% which is considered as the threshold for fault-tolerant quantum computation [11]. Figure 5**a** illustrates the *log-log* plot phase diagram of fidelity as a function of the $\gamma_\phi$ and $\kappa$.

Evidently, the decline of $\gamma_\phi$ and $\kappa$ is an efficient solution to increase QST fidelity. The white line in Fig. 5**a** signifies that when the $\gamma_\phi$ and $\kappa$ fall below this threshold, the QST fidelity can attain a remarkable 99%. The quadrangular star denotes the realistic metrics of $\gamma_\phi$ and $\kappa$ realized in one DQD-cavity hybridization device[17], indicating that the parameters of spin qubit and cavity need to be improved approximately five times in one hybridization device to surpass 99%. However, the requirements of $\gamma_\phi$ and $\kappa$ can be reduced by enhancing coupling strength ($g_c$). As shown in Fig. 5**b**, the fidelity phase diagram as a function of $g_c$ and $\kappa$, $\gamma_\phi$ is simulated ($\kappa = \gamma_\phi$ for simplification). As $g_c$ increases, the critical value of $\kappa$, $\gamma_\phi$ to breakthrough 99% is gradually raised (white line) and approached realistic parameters (dashed line in Fig. 5**b**). Theoretically, $g_c$ can be improved by increasing electrical dipole and lever arm in compact structure[16]. It has been achieved to be 500 MHz in Si-MOS[52] which makes the requirement of $\kappa$, $\gamma_\phi$ to realize 99% fidelity reduce to 0.35 MHz. Experimentally, the optimal $T_2^*$ of silicon spin qubit (black dashed line in Fig. 5**a**) and quality factor of high kinetic inductance superconducting cavity have been reached 20 μs[6] and $10^6$, which are already significantly superior to the above requirements. To keep the outstanding performance during integration, the application of on-chip filters[53] and flip-chips[54] can help. Moreover, the spin valve scheme itself can also be used to improve coherence, owing to synthetic SOC is switched off at most of time to isolate spin qubits from charge noise. However, we note that 99% resonant QST is challenging to reach in static SOC architecture even when the $\gamma_\phi$ and $\kappa$ are reduced by one order of magnitude (Fig. S7) due to the inevitable conflict between LZ transition and decay.

**Conclusion**

In summary, we propose a switchable SOC architecture for silicon spin qubits via spin valves by ultrafast electrically changing its magnetization configuration. Based on such architecture, we first confirm that the two-qubit SWAP operation of neighbor spins can be easily realized with fidelity exceeding 99%. Next, resonant QST between remote spin qubits is simulated. By optimizing the spin valve structure, pulse sequence and considering realistic parameters of spin qubit and cavity, the fidelity of QST also exceeds 99%, which is difficult to realize in the static SOC architecture. Moreover, our architecture also provides additional degree of freedom for spin qubit control via ultrafast modulating its spin-charge hybridization, which without disturbing the electron wavefunction to weaken the influence of crosstalk, chemical potential inhomogeneity and charge noise. Therefore, our results not only pave the way for construction of silicon quantum chips but also expand the research scope of spin valves.

## Methods

**Stray field simulation and optimization of spin valves**

The spin valve-induced 3D stray field distribution can be simulated by Mathematica RADIA package[55,56]. Depending on the orientation of magnetization of pinning/free layers in two spin valves, there are 16 types of stray field distributions in DQDs region. The synthetic SOC component from all the above 16 cases is determined by extracting the *x*-axis magnetic field gradient along z-direction ($b_{coup} = \frac{\partial B_x}{\partial z}$, Fig. S1 for detail) from 3D stray field. In main text, we only focus on the P-P and AP-AP cases to switch spin-photon coupling and magnetic field difference for QST and SWAP operation, where we choose the *x*-axis magnetic field difference between two QDs for simplification. To optimize the spin valve structure, we choose dephasing component of stray field as an evaluation standard by extracting the z-axis field gradient along z-direction ($b_{deph} = \frac{\partial B_z}{\partial z}$). By sequentially changing gap, size, thickness and shape of spin valves, the corresponding $b_{deph}$ is calculated to determine the optimal structures of spin valves (more detailed optimization process in Supplementary Note 1 and Fig. S2-3).

**Simulation of SWAP operation's fidelity**

To estimate the fidelity of SWAP operation, we numerically solve the master equation (Eq. 3 in main text) by considering the Heisenberg Hamiltonian:

$$H(t) = \Delta E_z \left(S_{z,1} - S_{z,2}\right)/2 + J(\boldsymbol{S}_1 \cdot \boldsymbol{S}_2 - 1/4),$$

where $\boldsymbol{S}_i = (S_{x,i}, S_{y,i}, S_{z,i})$ is the spin operator, $J$ represents the exchange coupling and $\Delta E_z$ is the Zeeman energy difference between two spin qubits which can be set to be zero in our spin valve scheme. The average fidelity is calculated by following equation:

$$\bar{F}(\mathcal{E}, U) = \frac{\sum_j \mathrm{tr}\left(UU_j^\dagger U^\dagger \mathcal{E}(U_j)\right) + d^2}{d^2(d+1)},$$

where $d$ is dimension of the state space. $U_j$ represents a group of unitary operators, $U$ is the ideal quantum gate and $\mathcal{E}$ is real quantum operation.

**Simulation and optimization of QST**

To explore the microwave photon-mediated QST between remote spin qubits in our spin valve scheme, the time-dependent part of $H_{sys}$ can be transformed into following form:

$$H(t) = B_j^Z \begin{pmatrix} \cos(\Phi_a) & 0 & 0 & \sin(\Phi_a) \\ 0 & -\cos(\Phi_b) & -\sin(\Phi_b) & 0 \\ 0 & -\sin(\Phi_b) & \cos(\Phi_b) & 0 \\ \sin(\Phi_a) & 0 & 0 & -\cos(\Phi_a) \end{pmatrix} - b_j^x \begin{pmatrix} -\sin(\Phi_a) & 0 & 0 & \cos(\Phi_a) \\ 0 & -\sin(\Phi_b) & \cos(\Phi_b) & 0 \\ 0 & \cos(\Phi_b) & \sin(\Phi_b) & 0 \\ \cos(\Phi_a) & 0 & 0 & \sin(\Phi_a) \end{pmatrix},$$

where the $B_j^Z$ and $b_j^x$ represent the z-axis stray field and x-axis stray field gradient at quantum dot $j$ and $\Phi_{a(b)} = B_x/(2t_c \pm B_z)$. Then, we can work on the $H_{DQD} + H_r$ eigen-basis and directly calculate the state evolution between spin qubits and photon. To optimize the QST, we first determine the compensatory time ($\delta T$) to recover the state leakage during switching processing. Then, the $t_{switch}$ and $\delta T$ are fixed in following optimization by changing the $\kappa$, $\gamma_\varphi$ and $g_c$ as shown in Fig. 5. Detailed simulation and optimization of QST for spin valve scheme and other methods are shown in Supplementary Note 3 and Note 4.

**Data availability**

The data that support the plots within this paper and other findings of this study are available from the corresponding authors upon request.

**Acknowledgements**

We thank Yong-Qiang Xu, Rui Wu, Ming Ni, Guang Yang, Kewen Shi, Le Zhao and Mengqi Zhao for helpful discussion. We acknowledge the financial support from National Natural Science Foundation of China (Grants No. 12304560, No. 92265113, No. 12074368, and No. 12034018), the Innovation Program for Quantum Science and Technology (Grant No. 238 2021ZD0302300) and China Postdoctoral Science Foundation (Grants BX20220281 and 2023M733408).

**Author contributions**

R. C. and G. C. conceived the project. F. L. and R. C. performed the theoretical simulation. B. W. and H. L. helped with SWAP simulation. R. C., F. L. and G. C. analyzed the data and wrote the manuscript. G. G. supervised the project. All the authors discussed the results and contributed to the final version of the manuscript.

**Competing Interests**

The authors declare no competing interests.

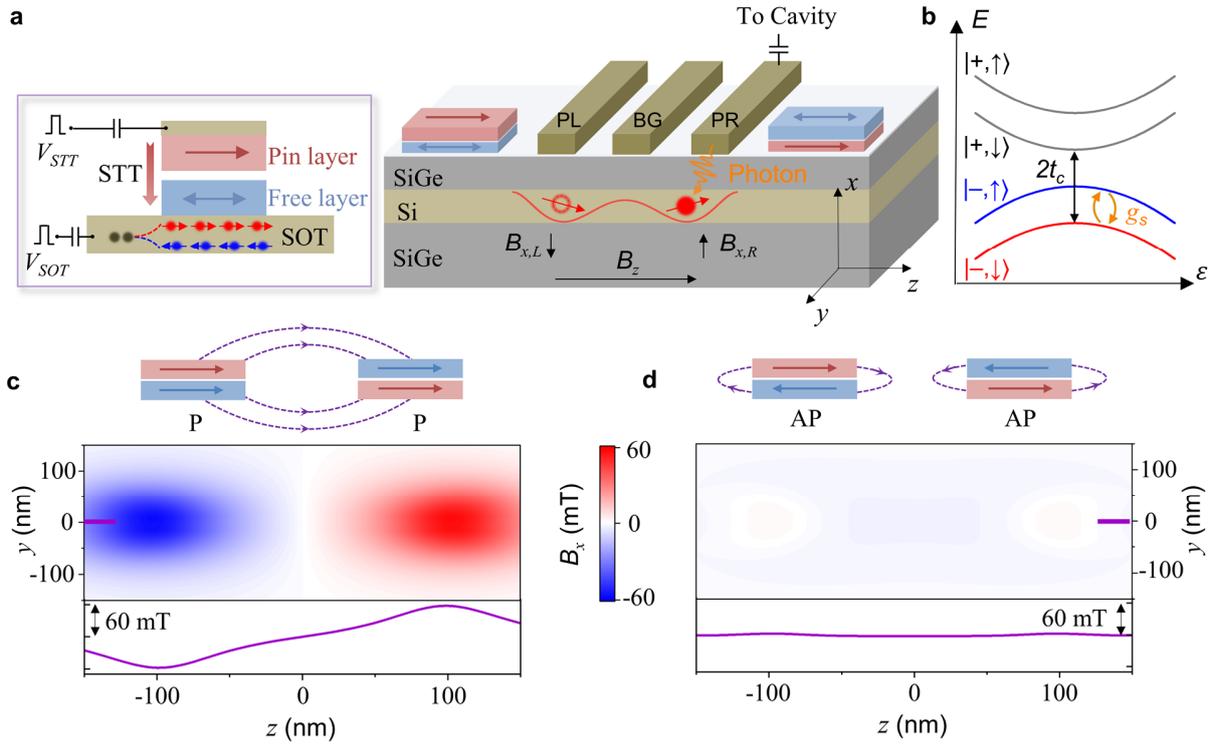

**Figure 1. Schematic of a switchable spin-orbit coupling based on spin-valves. a**. Schematic of a spin qubit in silicon double quantum dots (DQDs) accompanied by two spin valves to generate magnetic field gradient. Inset: schematic of a three terminal magnetic tunneling junction device for spin valve electrical switching via spin-transfer torque (STT) and spin-orbit torque (SOT). **b**. DQD energy level as a function of interdot energy detuning ($\varepsilon$). **c, d**. *x*-axis stray field distribution in the DQD region when two spin-valves are in parallel (P) or antiparallel (AP) states, respectively.

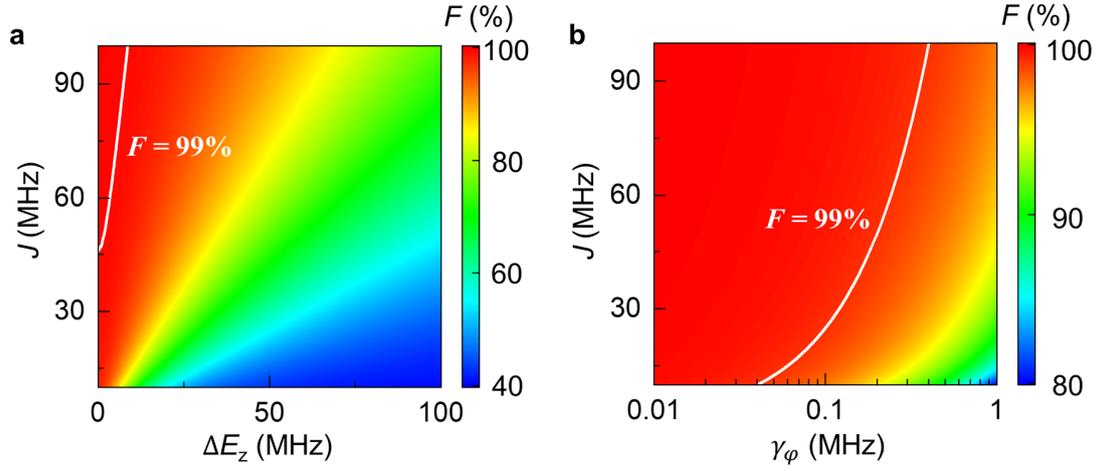

**Figure 2. SWAP operation between local spin qubits. a**. SWAP fidelity as a function of exchange coupling strength ($J$) and $\Delta E_z$ between silicon DQD for spin decoherence rate $\gamma_\varphi$ = 0.1 MHz. **b**. SWAP fidelity as a function of $J$ and $\gamma_\varphi$ for spin valve scheme with $\Delta E_z = 0$.

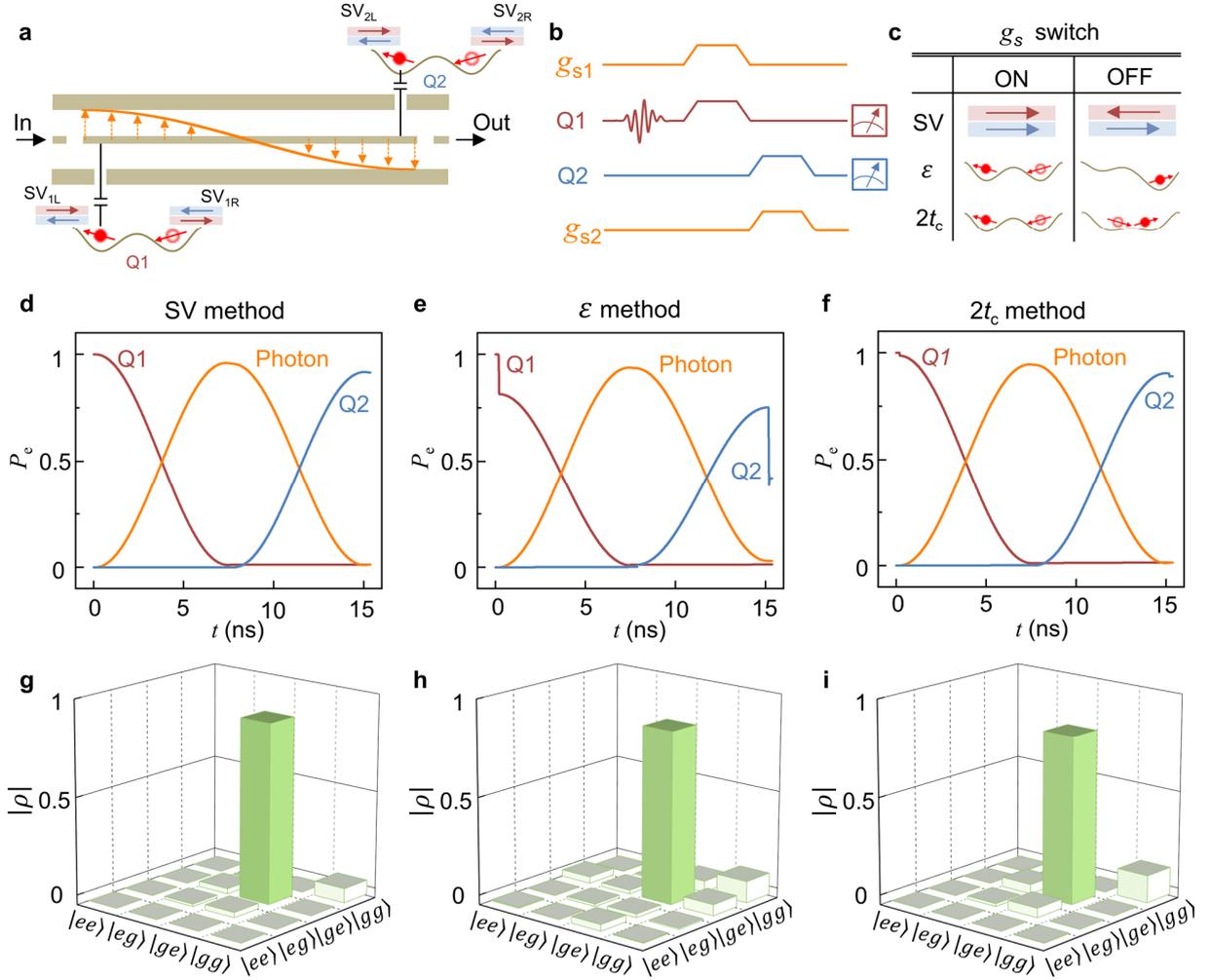

**Figure 3. Resonant quantum state transfer between remote spin qubits. a**. Schematic of two spin qubits in silicon DQDs are capacitively coupled to the end of a coplanar waveguide (CPW) superconducting resonator with spin valves to switch spin-photon coupling. **b**. Control pulse sequences for quantum state transfer (QST). **c**. Schematic of three switch methods for spin-photon coupling via spin valve (SV), interdot detuning ($\varepsilon$) or interdot tunnel coupling ($2t_c$). **d-e**. Quantum state evolution between qubit-1 (Q1), microwave photon and qubit-2 (Q2) during QST process controlled by SVs, $\varepsilon$ and $2t_c$, respectively. **g-i**. Reconstructed density matrix after QST for three control methods.

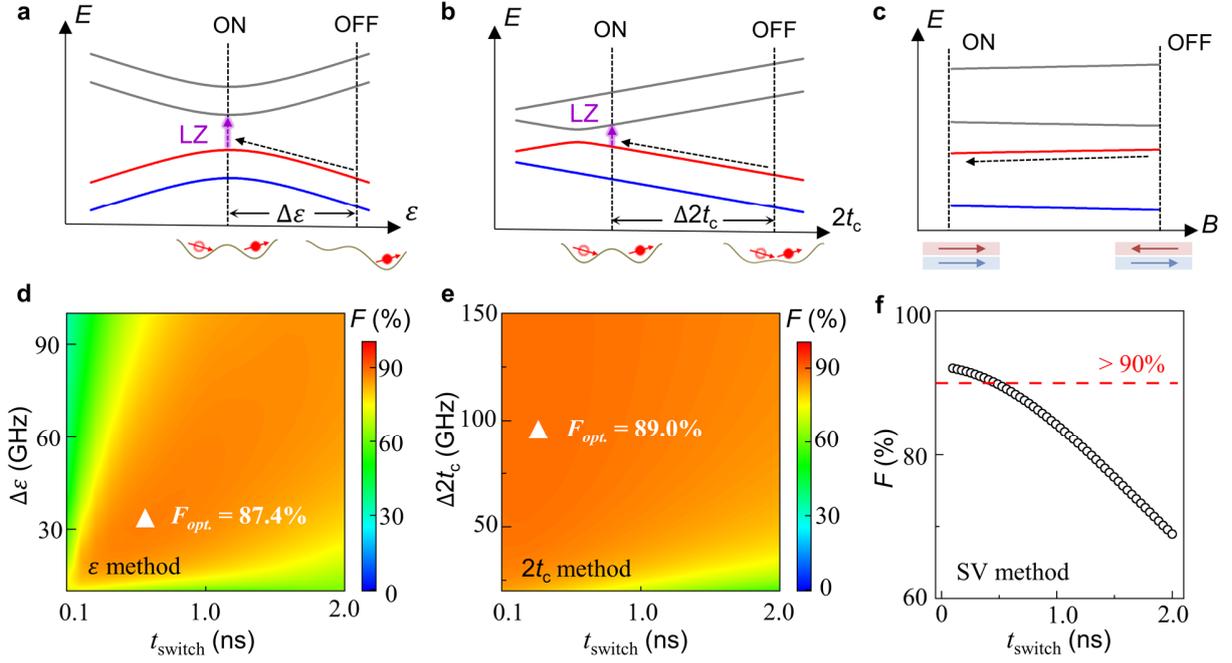

**Figure 4. Switching time dependent quantum state transfer fidelity. a-c**. Energy levels as a function of $\varepsilon$, $2t_c$ and magnetic field which correspond to three switching methods. **d**. Phase diagram of QST fidelity controlled by detuning method as a function detuning variation ($\Delta\varepsilon$) and switching time ($t_{switch}$). **e**. Phase diagram of QST fidelity controlled by tunnel coupling method as a function of tunnel coupling variation ($\Delta 2t_c$) and $t_{switch}$. The triangles in **d** and **e** represent optimal fidelity of this method. **f**. $t_{switch}$ dependent fidelity of spin-valve (SV) controlled QST.

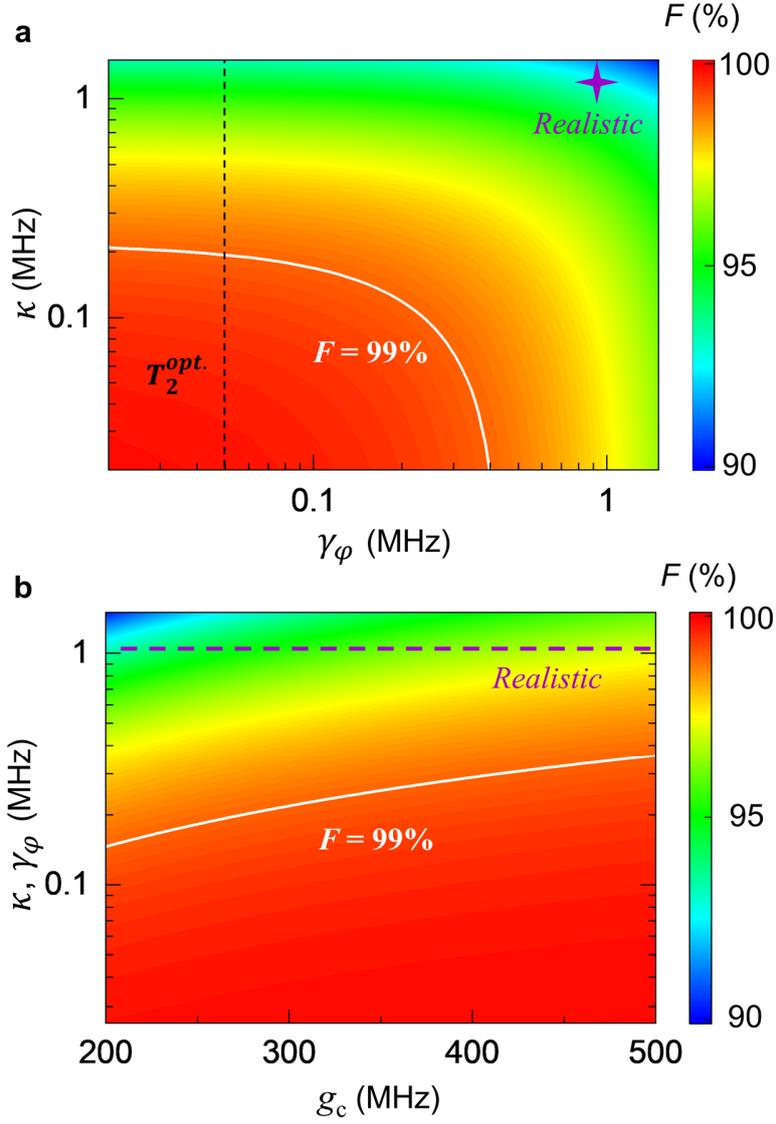

**Figure 5. Optimized resonant quantum state transfer. a**. Spin qubit dephase rate ($\gamma_\varphi$) and microwave photon loss rate ($\kappa$) dependent fidelity of QST controlled by spin valves. White solid line represents contour with fidelity equal to 99%. Black dashed line is the optimal dephase rate for silicon spin qubit with $T_2^{opt.} = 20$ μs[6]. **b**. DQD-resonant coupling strength dependent QST fidelity for spin valve method.